\documentclass[preprint, authoryear]{elsarticle}

\usepackage{natbib}
\usepackage[utf8]{inputenc} 
\usepackage[T1]{fontenc}
\usepackage{lmodern}
\usepackage[fleqn]{mathtools}
\usepackage{booktabs}
\usepackage{rotating}
\usepackage{amssymb}
\usepackage{amsmath}
\usepackage{graphicx}
\usepackage{caption}
\usepackage{epstopdf}
\usepackage{stackrel}
\usepackage{soul}
\usepackage{color}
\usepackage{float}
\usepackage[official]{eurosym}
\usepackage{tabularx}
\usepackage{multirow}
\usepackage{makecell}
\usepackage{lscape}
\usepackage{afterpage}
\usepackage{url}
\usepackage{mathrsfs}
\usepackage{chngcntr}
\usepackage{caption}
\usepackage{subcaption}
\usepackage{wasysym}
\usepackage{xfrac}
\usepackage[table]{xcolor}

\usepackage{ulem}
\usepackage{todonotes}

\newcommand{\Prob}{\text{Prob}}
\newcommand*\diff{\mathop{}\!\mathrm{d}}

\newdefinition{rmk}{Remark}
\newproof{pf}{Proof}
\newproof{pot}{Proof of Theorem \ref{thm2}}

\makeatletter
\def\ps@pprintTitle{%
	\let\@oddhead\@empty
	\let\@evenhead\@empty
	\def\@oddfoot{\centerline{\thepage}}%
	\let\@evenfoot\@oddfoot}
\makeatother

\begin{document}
\pagenumbering{gobble}
\title{On the Connection between Temperature and Volatility in Ideal Agent Systems \\[12pt]
% \title{On the Fundamental Problem of the Connection between 
% Temperature and Volatility in Econophysics \\[12pt]
{\small(First Draft: \today)}}
%\tnoteref{t1,t2}}
%\tnotetext[t1]{This document is a collaborative effort.}
%\tnotetext[t2]{The second title footnote which is a longer.}

\author[hhu]{Christoph J.\ Börner}
\ead{Christoph.Boerner@hhu.de}	
\author[hhu]{Ingo Hoffmann\corref{cor1}}
\ead{Ingo.Hoffmann@hhu.de}	
\author[hhu]{John H.\ Stiebel}
\ead{John.Stiebel@hhu.de}

\cortext[cor1]{Corresponding author. Tel.: +49 211 81-15258; Fax.: +49 211 81-15316}
%\cortext[cor2]{Principal corresponding author}
%\fntext[fn1]{This is the specimen author footnote.}
%\fntext[fn2]{Another author footnote, but a little more longer.}
%\fntext[fn3]{Another author footnote, but a little more longer.}

\address[hhu]{Financial Services, Faculty of Business Administration and Economics, \\ Heinrich Heine University D\"usseldorf, 40225 D\"usseldorf,
	Germany}	

\begin{abstract}
Models for spin systems known from statistical physics are applied by
analogy in econometrics in the form of agent-based models. Researchers
suggest that the state variable temperature $T$ corresponds to
volatility $\sigma$ in capital market theory problems. To the best of
our knowledge, this has not yet been theoretically derived, for example, for an ideal agent system. 
In the present paper, we derive the exact algebraic relation between $T$ and $\sigma$ for an ideal agent system and discuss implications and limitations.
\end{abstract}

\begin{keyword}
	Agent System
	\sep Econophysics
	\sep Temperature
	\sep Volatility \\[6pt] 
	
	
	\noindent \textit{ORCID IDs:} 
	0000-0001-5722-3086 (Christoph J.~B\"orner), 
	0000-0001-7575-5537 (Ingo Hoffmann), 
	0000-0003-0088-2456 (John H.~Stiebel),    	\\[6pt] %Geprüft: IHo, 10.11.2021
	
	%\noindent \textit{Acknowledgement:} We thank $\ldots$. 
\end{keyword}
\maketitle
\newpage

%\pagenumbering{arabic} \setcounter{page}{1}
\pagenumbering{arabic}

\section{Introduction} \label{Introduction}
In the context of econophysics, methods from statistical physics are used to model the behavior of investors (so-called agents), for example, to draw conclusions about the price movement of a stock and arrive at a one-step ahead forecast model; see, e.g., \citet{Vikram.2011}. Starting with the work of \citet{Weidlich.1971} and \citet{Galam.1982}, the research field of econophysics utilizing spin systems has steadily developed and branched out; see, e.g., \citet{Bouchaud.2013, Sornette.2014} for a review.
The models of econophysics, based on the analogy with spin systems from statistical physics, depend on a state variable $T$ that describes the system and is called temperature in statistical physics.
For temperature in statistical physics, see, e.g., \citet{Isihara.1971, Landau.1980, Greiner.1995, Kardar.2007}.
In econophysics, the state variable $T$ is referred to differently in
various applications: noise, irrationality, degree of randomness in
agents' decisions, the collective climate parameter or volatility; see, e.g., \citet{Weidlich.1971, Kaizoji.2000, Kozuki.2003, Oh.2007, Kleinert.2007, Kozaki.2008, Krause.2012, Bouchaud.2013, Crescimanna.2016}.
In models for capital markets in which temperature $T$ is identified with volatility $\sigma$, e.g., of a stock, various relationships are assumed: 
$T \propto \sigma^2$ \citep{Kozuki.2003}, 
$T = \tfrac{\sigma}{\sqrt{2}}$ \citep{Kleinert.2007}, 
$T \propto \sigma$ \citep{MattosNeto.2011}.

To the best of our knowledge, the relationship $T \propto \sigma $ is predominantly used in capital market theory applications \citep{MattosNeto.2011, Bouchaud.2013, Boerner.2022}, 
but the exact algebraic relation does not appear to have been theoretically derived, even for an ideal agent system. 
This contribution is dedicated to this precise question and is intended to close the research gap reflected in the variety of algebraic equations adopted and proposed \citep{Kozuki.2003, Kleinert.2007, MattosNeto.2011}.
The function $f\colon \sigma \rightarrow T$ is derived from theory. Implications are examined, and the limits that need to be considered are identified. For the simple, ideally designed example of a two-state agent system \citep{Bouchaud.2013} the equation $T = f(\sigma) $ is derived and further analyzed.

This study is structured as follows. In the next section, temperature is defined as state variable $T$. In Section \ref{AgentSystems}, the example of a two-state ideal agent system is introduced. Section \ref{StochModellMarketCap} is devoted to determining the central equation for deriving measurement equations. An application to the ideal agent system and a conclusion follow.

\section{Definition of $T$} \label{Temperaturedefinition}
In physics, since the revision of the international system of units, temperature $T$ has been linked to thermal energy using the Boltzmann constant $k$ 
\citep[p.\ 133]{InternationalBureauofWeightsandMeasure.2019}. 
If the internal energy $E = E(S, {\bf X})$ is described as a function of the entropy $S$ and possibly other state variables $\bf X$ and the entropy is calculated from the microcanonical partition function $\Omega$ using the equation $S = k \ln\Omega$, then 
\begin{flalign} \label{StatPhysTemp}
	T :=  \left. \frac{\partial E}{\partial S}\right|_{{\bf X} = {\rm const.}}
\end{flalign}
is the definition of temperature in statistical physics; see, e.g., \citet{Greiner.1995}.

With the transformation $E= - U$, \citet{Marsili.1999} connects the internal energy $E$ with economic utility $U$ and arrives at the definition of the state variable $T$ in econometrics:
\begin{flalign} \label{ÖkoTemp}
	T :=  - \left. \frac{\partial U}{\partial S}\right|_{{\bf X} = {\rm const.}}
\end{flalign}
Where $\bf X$ summarizes other variables, for example, the message environment $\cal B$ in the ideal agent system \citep{Bouchaud.2013, Boerner.2022}.

If finite changes are considered with constant ${\bf X}$, then Equation (\ref{ÖkoTemp}) leads to the simple relation $\Delta U  \propto - T \Delta { S}$.
That is, with fixed $T$ and increasing entropy, utility decreases. \citet[p.\ 13]{Marsili.1999} concluded that state variable $T$ "[$\ldots$] {\it can be considered as the price of (negative) entropy}".

If and only if the definition of Equation (\ref{ÖkoTemp}) is used in econophysics, a consistent theory can be constructed based on the methods of statistical physics. The variable $T$ is initially used as a parameter in all equations and must be identified based on the specific circumstances of the application under consideration. This is specifically shown for the example of a two-state ideal agent system.

\section{Two-State Ideal Agent System}\label{AgentSystems}
A number $N$ of stock investors will be considered as an example of a system of agents. A news situation $-1\leq {\cal B} \leq +1$ is given and influences the investors. For the sake of simplicity, it is assumed that each investor only acts based on the news situation ${\cal B}$ and only buys or sells one share at a time. Agreements or alliances between investors are excluded, so each investor acts in isolation and is uninfluenced by other investors. These systems can be referred to as systems without interactions or ideal agent systems; see, e.g., \citet{Bouchaud.2013, Boerner.2022}.

In the following, w.l.o.g., a positive, constant news situation with
strength $|{\cal B}| = B$ is assumed. Investors can act in
"conformity" to the news, i.e.,\ buy if the news is positive, or in "non-conformity", i.e.,\ sell if the news is positive. The state of an investor $i = 1, \ldots, N$ is described with $s_i = +1$ (conform) or $s_i = -1$ (non-conform).
If a dynamic equilibrium is established, then there is a number $N_{+}$ of "conform" and a number $N_{-}$ of "non-conform" investors, with $N = N_{+} + N_{-}$. If an inversion of the agent system is ruled out, i.e.,\ the overall system acts in accordance with the positive news, the difference $ N_{+} - N_{-}$ is positive and corresponds to a buyer surplus.
In econometric applications, the normalized variable $M = \frac{1}{N} (N_{+} - N_{-}) = \frac{1}{N} \sum_i s_i $ can be referred to as trade potential ({\it phys.}: $\propto$ magnetization), see, e.g., \citet{Boerner.2022}, and can be used, e.g., for one-step ahead forecast models \citep{Vikram.2011}.

\subsection{Occupation Probabilities} \label{OccupationProb}
For the two states, occupation probabilities can be specified with the help of the logit rule \citep{Bouchaud.2013}, which corresponds to the Boltzmann-Gibbs distribution in statistical physics \citep{Landau.1980, Greiner.1995}.

Let $x = \frac{T_0}{T}$, and $T_0$ is calculated from all constant system parameters and constant external conditions (e.g.,\ the constant news $B$); see, e.g., \citet{Bouchaud.2013, Boerner.2022}. Then, the probabilities for each state $s = (-1, +1)$ of the agent are:
\begin{flalign}\label{Probabilities} \nonumber
	P_{-} = \Prob(s = -1) 
	& = \frac{\exp(-x)}{\exp(-x) +  \exp(+x)}\\ 
	\\[-12pt] \nonumber
	P_{+} = \Prob(s = +1) 
	& = \frac{\exp(+x)}{\exp(-x)  + \exp(+x)}
\end{flalign}
with $P_{-} + P_{+} = 1$. A similar representation of the occupation
probabilities can be found in, e.g., \citet{Bouchaud.2013}. 
Equation (\ref{Probabilities}) can be derived within the framework of econophysics and is consistent with the definition given in Equation (\ref{ÖkoTemp}).
Accordingly, $T$ is initially only a parameter, and the question of whether $T$ is related to the volatility of the stock remains open.

\subsection{Distribution of the Trade Potential $M$} \label{DistributionM}
For a finite number $N < \infty$ of isolated investors (ideal agents),
the occupation numbers $ N_{-}$ and $ N_{+}$ are stochastically
dependent random variables, and thus the trade potential $M$ is also
a random variable. If all investors are lined up for illustration, it
becomes clear that both $ N_{-}$ and $ N_{+}$ follow a binomial
distribution: ${\rm Binom}(N_{-}; N, P_{-})$ and ${\rm Binom}(N_{+};
N, P_{+})$; for a definition, see, e.g., \citet{Abramowitz.2014}.
This allows the expected value $\mu_M = P_{+} - P_{-}$ and the variance $\sigma_M^2 = \frac{4}{N} P_{+}P_{-}$ to be calculated for the random variable $M$.
The derivation is almost identical to the calculation in \citet[Chap.\
2]{Fliebach.2018}, with the difference being that the normalized variable $M$ is considered here.
For large $N$, the above binomial distributions can be approximated by normal distributions \citep{Kendall.1977}. 
This means that the distribution of the random variable $M$ can also be described by a normal distribution: $M \sim {\cal N} (\mu_M, \sigma_M)$.
The variance is proportional to $\frac{1}{N}$, so for large $N$, the relative width of the distribution tends to zero and is thus sharply localized around $\mu_M$, i.e.,\ fluctuations in $M$ are almost no longer observed; see \citet{Fliebach.2018}, and also compare \citet{Greiner.1995}.

\section{Stochastic Model of Market Capitalization}\label{StochModellMarketCap}
Let the news situation ${\cal B} > 0$ be constant for a finite period $\mathbb{T}$. If this period is broken down into finite subperiods $\Delta t$, then in each subperiod $[t, t+\Delta t]$ there is a number $N_{+}$ of buyers and a number $N_{-}$ of sellers. At the beginning of the period, the last quoted price of the stock $p_t$ is known, and the potential rate of change in market capitalization ($V_t = N p_t$) that will take effect at the end of the subperiod can be inferred:
\begin{flalign}\label{MarketCap_1} \nonumber
	\Delta V_t 
	& = p_t (N_{+} -  N_{-})  \Delta t \\
	& = V_t M_t  \Delta t
\end{flalign}
with $M_t \sim {\cal N} (\mu_M, \sigma_M)$ in each subperiod. Thus,
the change $\Delta V_t$ is a random variable and changes in each subperiod. The time-discrete representation of a stochastic process becomes observable. Such processes are described by stochastic differential equations; see, e.g., \citet{Hull.2018}. Following technics described in \citet[Chap.\ 3]{Wilmott.1998} and \citet[Chap.\ 14]{Hull.2018}, a stochastic differential equation with drift term ($\propto \mu_M$) and diffusion term ($\propto \sigma_M$) for the stochastic process can be written:
\begin{flalign}\label{StochProzessMC} 
	\diff V_t 
	& = V_t \mu_M  \diff t + V_t \sigma_M  \diff W_t,
\end{flalign}
where $W_t$ denotes a Wiener process. The transformation $X_t = G(V_t,
t) = \ln(V_t) - \ln(N)$ transforms to the logarithmic price $X_t$ of
the stock. Using It\={o}'s lemma with the partial derivatives
$\partial_V G = \frac{1}{V}$, $\partial_{VV} G = -\frac{1}{V^2}$ and
$\partial_t G = 0$, it follows that:
\begin{flalign}\label{StochProzessMC_2} 
	\diff X_t 
	& =\left( \mu_M - \frac{\sigma_M^2}{2} \right) \diff t + \sigma_M  \diff W_t.
\end{flalign}
Only one price is quoted on the capital market at a time. This means that there is only one logarithmic price at a time. The stochastic process for logarithmic returns \citep[Eq.\ 14.17]{Hull.2018} and the process defined with Equation (\ref{StochProzessMC_2}) must therefore be identical. 
A comparison of coefficients then provides the following relationship:
\begin{flalign}\label{MainResults} 
	\mu_X = \mu_M & \quad {\rm and} \quad \sigma_X = \sigma_M.
\end{flalign}
Where $\mu_X$ is the expected value of the logarithmic returns and $\sigma_X$ is the standard deviation. The latter is therefore the volatility of the stock under consideration.

Note that it was assumed that the positive news situation ${\cal B}$
is constant for the period $\mathbb{T}$. In practice, this will only
happen for a short period (a few minutes) of a daily trading
session. The relationships described in Equation (\ref{MainResults})
should therefore only be observable in practice for short periods of
time. If the focus is on one-step-ahead forecasts, mean values and
sample variances calculated over short periods of time are to be
preferred as estimators for $\mu$ and $\sigma$. Furthermore, note that Equation (\ref{StochProzessMC_2}) does not capture contributions to expected value and volatility due to changes in news flow ${\cal B}$.

Section \ref{DistributionM} showed that $\sigma_M \propto \frac{1}{\sqrt{N}}$. With Equation (\ref{MainResults}), this also applies to $\sigma_X$ for short periods of time with a constant news situation. The latter means that the volatility that can be observed over short periods of time decreases as the number of investors $N$ increases and in the limiting case of an infinite number of investors approaches zero.
A similar phenomenon was described by \citet{Bouchaud.2013} in the
context of socioeconomic issues. There, however, it related to noise levels that influence decision-making situations and scale with $\frac{1}{\sqrt{N}}$ depending on the population size.

Equation (\ref{MainResults}) provides a very general connection between the capital market quantities and the distribution parameters of the trade potential $M$. However, $|M| \leq 1$ is limited and thus also $\sigma_M$. The validity of $\sigma_X = \sigma_M$ must always be checked for large volatilities $\sigma_X$, in particular whether strongly fluctuating messages or high-frequency message changes could be the reason for the high volatility.

\section{Application}\label{Applicatiton}
In the following, the central equation $\sigma_X = \sigma_M$ for the example of the two-state ideal agent system from Section \ref{AgentSystems} is used to establish the connection between $T$ and the volatility $\sigma_X$:
\begin{flalign}\label{TempVola} \nonumber
	\sigma_X  & = \sigma_M \\ \nonumber
		      & = \sqrt{\frac{4}{N} P_{+}P_{-}} \\ \nonumber
	    	  & = \sqrt{\frac{4}{N} \frac{1}{4\cosh^2(x)}} \\
	          & = \frac{1}{\sqrt{N}} \, {\rm sech} (x).
\end{flalign}
With $x = \frac{T_0}{T}$, the measurement equation follows:
\begin{flalign}\label{MeasurementEq} 
	T    & = \frac{T_0}{{\rm sech}^{-1} (\sqrt{N} \sigma_X)}.
\end{flalign}
Where ${\rm sech}^{-1}(u)$ is the inverse hyperbolic secant function, and $\sigma_X$ as the volatility of the logarithmic returns is now the observable that can be measured on the capital market. Equation (\ref{MeasurementEq}) indirectly measures the state variable $T$. For the two-state ideal agent system, it is the state variable that is consistent according to Equation (\ref{ÖkoTemp}). In further state equations for the system, this state variable determined in this way must be used to maintain consistency within the theory.

\section{Conclusion}\label{Conclusion}
The highly simplified example of a two-state ideal agent system was
studied in detail, and the relationship between the volatility
$\sigma_X$ and the state variable $T$ was established. Even this
simplest conceivable example shows that the relationship between
$\sigma_X$ and $T$ can be nonlinear. It can be assumed that this is
also the case in other applications. This consideration is not
exhaustive; Equation (\ref{MeasurementEq}) is a special measurement
equation for the example, and in individual cases, the measurement equation $T = f(\sigma_X)$ must always be derived. 
This allows for the indirect and theory-conforming determination of the state variable $T$ from the measured volatility $\sigma_X$. If a linear relationship $T = a \sigma_X +b$ is required, then the measurement equation can be linearized around an operating point $\sigma_X^0$, i.e.,\ considering a Taylor series up to the linear term. In practice, questions about the validity of the approximation (keyword: measuring range) then have to be answered.

%\newpage
%
%\section*{References}

\bibliography{../../020_Literatur/005_CitaviBibTexFile/CitaviHerding}

\end{document}